\documentclass[twocolumn,showpacs,preprintnumbers,amsmath,amssymb,floatfix]{revtex4}

\usepackage{graphicx}
\usepackage{dcolumn}
\usepackage{bm}

\begin{document}

%

\let\a=\alpha      \let\b=\beta       \let\c=\chi        \let\d=\delta
\let\e=\varepsilon \let\f=\varphi     \let\g=\gamma      \let\h=\eta
\let\k=\kappa      \let\l=\lambda     \let\m=\mu
\let\o=\omega      \let\r=\varrho     \let\s=\sigma
\let\t=\tau        \let\th=\vartheta  \let\y=\upsilon    \let\x=\xi
\let\z=\zeta       \let\io=\iota      \let\vp=\varpi     \let\ro=\rho
\let\ph=\phi       \let\ep=\epsilon   \let\te=\theta
\let\n=\nu
\let\D=\Delta   \let\F=\Phi    \let\G=\Gamma  \let\L=\Lambda
\let\O=\Omega   \let\P=\Pi     \let\Ps=\Psi   \let\Si=\Sigma
\let\Th=\Theta  \let\X=\Xi     \let\Y=\Upsilon

%

%

\def\cA{{\cal A}}                \def\cB{{\cal B}}
\def\cC{{\cal C}}                \def\cD{{\cal D}}
\def\cE{{\cal E}}                \def\cF{{\cal F}}
\def\cG{{\cal G}}                \def\cH{{\cal H}}
\def\cI{{\cal I}}                \def\cJ{{\cal J}}
\def\cK{{\cal K}}                \def\cL{{\cal L}}
\def\cM{{\cal M}}                \def\cN{{\cal N}}
\def\cO{{\cal O}}                \def\cP{{\cal P}}
\def\cQ{{\cal Q}}                \def\cR{{\cal R}}
\def\cS{{\cal S}}                \def\cT{{\cal T}}
\def\cU{{\cal U}}                \def\cV{{\cal V}}
\def\cW{{\cal W}}                \def\cX{{\cal X}}
\def\cY{{\cal Y}}                \def\cZ{{\cal Z}}

%

\newcommand{\Ns}{N\hspace{-4.7mm}\not\hspace{2.7mm}}
\newcommand{\qs}{q\hspace{-3.7mm}\not\hspace{3.4mm}}
\newcommand{\ps}{p\hspace{-3.3mm}\not\hspace{1.2mm}}
\newcommand{\ks}{k\hspace{-3.3mm}\not\hspace{1.2mm}}
\newcommand{\des}{\partial\hspace{-4.mm}\not\hspace{2.5mm}}
\newcommand{\desco}{D\hspace{-4mm}\not\hspace{2mm}}

\def\deltaakpi{\Delta {{A}}_{K\pi}}


\title{\boldmath 
 CP asymmetry in $B^{+}\to K^+\pi^{0}$ and New Physics }

\author{Namit Mahajan
}
\email{nmahajan@prl.res.in}
\affiliation{
 Theoretical Physics Division, Physical Research Laboratory, Navrangpura, Ahmedabad
380 009, India
}


\begin{abstract}
The CP asymmetry in $B^0\to K^+\pi^-$ is expected to be similar to
that in $B^+\to K^+\pi^0$. The experimental data however show $\sim 5\sigma$
difference between the two, leading to the so called $\Delta {{A}}_{K\pi}$
puzzle. Employing sum rule(s) following from (approximate) flavour symmetry, we show that it is possible to accommodate the observed experimental values within the standard model (SM) for a narrow range of parameters. Sub-leading terms can bring the theoretical predictions in better agreement with the data. Resolution via modified electroweak penguin contributions is possible for a large CP violating phase generated by the new physics. However, the data on polarization in $B\to VV(T)$, $B_s$-$\bar{B}_s$ mixing (and large CP phase) and $B^+\to\tau^+\nu_{\tau}$ rate can not be simultaneously accommodated within SM or new physics with only enhanced electroweak penguins. A plausible resolution to these, and not spoiling the $B\to K\pi$ rates and asymmetries, could be a general two Higgs doublet model.

\end{abstract}

\maketitle


Rare B-decays provide a golden opportunity to test the fundamental interactions
among the elementary particles, and at the same time allow us to extract valuable
information about the dynamics of strong interactions. The study of CP violation,
and its origin, has been one of the main aims of the B-factories. Thanks to the excellent
experimental precision reached at the B-factories, and also at CLEO and Tevatron,
we now have accurate measurements of branching ratios and CP asymmetries for many
rare decay processes. This on one hand has established the dominance of the phase
in the CKM matrix as source of CP violation (at least at low energies), while on 
the other hand has brought some tantalizing issues in light, possibly providing the 
first glimpse of physics beyond the standard model (SM). On the theoretical front,
accurate experimental measurements have pushed the calculations to a higher level,
in some case demanding very accurate estimation of sub-leading effects and theoretical
errors. SM, by large, still remains highly successful in explaining the data. Any
attempt to infer hints of new physics (NP) need to ensure that we have quantitatively
exhausted all the possibilities within SM, including sub-leading effects and any other neglected 
contributions based on some assumptions.

$B\to K\pi$ decays are dominated by $b\to s$ penguin transitions at the quark level. The
four branching ratios, four direct CP asymmetries and the indirect CP asymmetry in the
$B^0\to K_S\pi^0$ channel provide a wealth of information which can be used to extract weak phases
to enable us to reconstruct the unitarity triangle. In achieving this feat, some extra input
is needed from other well measured quantities. Table 1 summarises the present experimental situation
on $B\to K\pi$ modes. The entries correspond to the values obtained by the
Heavy Flavor Averaging Group \cite{HFAG}.

\begin{table}
\begin{center}
\begin{tabular}{|l|l|}
\hline
Observable & HFAG average \\ \hline 
$BR(B^0\to K^+\pi^-)$ & $(19.4\pm 0.6) \times 10^{-6}$ \\ 
$BR(B^0\to K^0\pi^0)$ & $(9.8\pm 0.6) \times 10^{-6}$ \\ 
$BR(B^+\to K^+\pi^0)$ & $(12.9\pm 0.6) \times 10^{-6}$ \\
$BR(B^+\to K^0\pi^+)$ & $(23.1\pm 1.0) \times 10^{-6}$ \\ 
$A_{CP}(K^+\pi^-)$ & $(-9.5\pm 1.2) \%$ \\ 
$A_{CP}(K^+\pi^0)$ & $(5\pm 2.5) \%$ \\ 
$A_{CP}(K^0\pi^+)$ & $(0.9\pm 2.5) \%$ \\ 
$C_{K_S\pi^0}$ & $0.01\pm 0.10$ \\ 
$S_{K_S\pi^0}$ & $0.57 \pm 0.17$\\ \hline
\end{tabular}
\caption{HFAG values \cite{HFAG} for observables in $B\to K\pi$ system}
\end{center}
\end{table}

The difference in direct CP asymmetries,
 $\Delta {{A}}_{K\pi}\equiv A_{CP}(K^+\pi^0) - 
A_{CP}(K^+\pi^-)$, has attracted lot of attention. Recently the Belle Collaboration
published \cite{bellenature} an updated and precise measurement of this quantity,
which is at variance with SM expectation.
From the table one can read off that
$\Delta {{A}}_{K\pi}=14.8\pm 2.8 \neq 0$ at $\sim 5\sigma$. 
A non-zero value of $\Delta {{A}}_{K\pi}$ has been 
argued to be a signature of new physics \cite{kpinp}, since the two amplitudes differ by terms which are small in magnitude. In order to infer that this is in fact a clear signal of NP or that its resolution necessarily calls for NP, it is
mandatory to carefully examine all the assumptions generally made in estimating
various contributions, and to estimate the impact of the neglected small terms in 
the analysis. Needless to say that parametric uncertainties need to be kept in mind
while making all these estimates. 

There are two approaches to non-leptonic B-decays. One is based on the effective
theory language and employs operators and short distance coefficients in the low energy
effective theory to describe the quark level decays (see for example \cite{buchallaburas}).
 The second is the diagrammatic approach
combined with flavour symmetries \cite{gronauetal}. 
In the present note, combining the available experimental information and theoretical constraints
with the flavour symmetries, we revisit the sum rule for CP asymmetry
in $B^+\to K^+\pi^0$ proposed in \cite{kpluspi0sumrule}. 
We invert the sum rule and examine to what extent 
the inverted relation, expressing a certain combination of theoretical
parameters, called $\delta_{EW}$ below, in terms of measurable quantities is satisfied within SM. 
We find that within the
present errors, the observed value for $\delta_{EW}$ does agree with the
SM prediction for the same quantity. We hasten to mention that the biggest uncertainty stems
from $\vert V_{ub}\vert$, and therefore requires a better precision on the same to conclude
anything about the presence of NP. 

The effective Hamiltonian responsible for the non-leptonic $b\to s$ transitions is given by \cite{buchallaburas}
($b\to d$ transitions are described by appropriate changes)
\begin{equation}
{\mathcal{H}}_{eff} = \frac{G_F}{\sqrt{2}}\left[\sum_{i=1,2} C_i(\lambda_u Q_i^u + \lambda_c Q_i^c)
- \lambda_t\sum_{i=3}^{10} C_i Q_i\right] + H.C \label{heffective}
\end{equation}
where $\lambda_q=V_{qb}^*V_{qs}$, $C_i$ are the relevant Wilson coefficients while $Q_i$ are four fermion
operators. Here, $Q_{1,2}^{u,c}$ are the current-current operators, while $Q_{3-6}$ and $Q_{7-10}$ are
the QCD penguin and electroweak (EW) penguin operators. Operators $Q_5$, $Q_6$, $Q_7$ and $Q_8$ have
$(V-A)\otimes (V+A)$ structure while all others have $(V-A)\otimes (V-A)$ structure. 
The NLO SM Wilson coefficients at scale $\mu=m_b$ (approximately) read:
\[
C_1\sim -0.3,\,\,\, C_2 \sim 1.14,\,\,\,  C_{3-6} \sim {\mathcal{O}}(10^{-2}),\]
\[ C_{7,8}\sim {\mathcal{O}}(10^{-4}),
\,\,\, C_9 \sim -1.28\alpha,\,\,\, C_{10}\sim 0.33\alpha
\]
Due to the smallness of $C_{7,8}$, the corresponding contributions can be safely 
neglected. This also holds for
extensions of SM where the $(V-A)\otimes (V+A)$ operators are not tremendously enhanced. Further,
$O_9$ and $O_{10}$ can be Fierz transformed into $O_1$ and $O_2$. 

Various $B\to K\pi$ decay amplitudes are expressed as \cite{gronauetal, gronau-rosner, yoshikawa}
\begin{eqnarray}
 A^{+0} &\equiv& \sqrt{2}A(B^+\to K^+\pi^0) = -(p + t + c + a) \nonumber \\
A^{0+} &\equiv& A(B^+\to K^0\pi^+) = (p + a)\nonumber \\
 A^{+-} &\equiv& A(B^0\to K^+\pi^-) = -(p + t) \\ \label{amplitudes}
 A^{00} &\equiv& \sqrt{2}A(B^0\to K^0\pi^0) = (p-c)\nonumber 
\end{eqnarray}
The amplitudes $p$, $t$, $c$ and $a$ are linear combinations of graphical amplitudes denoting
Tree ($T$), Colour suppressed Tree ($C$), QCD-Penguin ($P$), colour allowed and suppressed EW-Penguin  
($P_{EW}$ and $P_{EW}^C$), Annihilation ($A$), W-Exchange ($E$) and Penguin
Annihilation ($PA$) diagrams, expected to follow the hierarchy \cite{gronauetal}:
$\vert P\vert >> \vert T\vert \sim \vert P_{EW}\vert >> \vert C\vert \sim \vert P_{EW}^C\vert >
\vert A,E,PA\vert$. 
Diagrams $A$, $E$ and $PA$ involve the spectator quark and are therefore
expected to be suppressed ($\propto f_B/m_B$) and are often neglected. 
This expectation may not hold true in 
the presence of significant rescattering. Within the $B\to K\pi$ system, a large CP asymmetry
in $B^+\to K^0\pi^+$, $A_{CP}(K^0\pi^+)$, would indicate rescattering. From the table,
it is clear that this is not the case here and therefore amplitude $a$ can be safely neglected.
In such a case, we have $p = P - P_{EW}^C/3$, $t = T + P_{EW}^C$ and $c = C + P_{EW}$, where 
the u-quark penguin contribution has been absorbed in the definition of $T$ and $C$.
Further, it is implied that the penguin amplitudes $P,\, P_{EW},\, P_{EW}^C$ contains the CKM factor $V_{tb}^*V_{ts}$ while all others 
contain $V_{ub}^*V_{us}$. Using the unitarity of the CKM matrix, one can eliminate $V_{tb}^*V_{ts}$
in favour of $V_{cb}^*V_{cs}$ and $V_{ub}^*V_{us}$. This has the advantage of employing the
experimentally measured elements of the CKM matrix.
Another advantage is that in this form, generalization to include NP effects due to modified
penguin operators is straightforward - the modified coefficient can be simply made complex
to take care of the extra phases, if needed. 
 From the above expressions, one expects 
$A_{CP}(K^+\pi^-)\sim A_{CP}(K^+\pi^0)$ since the two amplitudes differ by small contributions.
The data however point to the contrary ($\Delta {{A}}_{K\pi} \neq 0$ at $\sim 5\sigma$). 
Neglecting small contributions,
following ratios of CP averaged rates are expected to be
(almost) unity within SM:
\begin{eqnarray}
 R &\equiv& \frac{\Gamma_{av}(B^0\to K^+\pi^-)}{\Gamma_{av}(B^+\to K^0\pi^+)} 
 \nonumber \\
 R_c &\equiv& \frac{2 \Gamma_{av}(B^+\to K^+\pi^0)}{\Gamma_{av}(B^+\to K^0\pi^+)} 
 \\ \label{rateratios}
 R_n &\equiv& \frac{\Gamma_{av}(B^0\to K^+\pi^-)}{2\Gamma_{av}(B^0\to K^0\pi^0)} 
 \nonumber 
\end{eqnarray}
The data in the table do follow this expectation. 
Using flavour $SU(3)$ symmetry, it is possible to relate the EW penguin contributions
to the tree contributions. Neglecting $C_{7,8}$,
and Fierz transforming $O_{9,10}$ immediately enables one to express the relevant terms
in the effective Hamiltonian such that \cite{deltaew, kpluspi0sumrule, neubert}
\begin{equation}
 t + c = T + C + P{EW} + P_{EW}^C = (T+C)[\delta_{EW} - e^{-i\gamma}] \label{tplusc}
\end{equation}
where $\gamma$ is the CKM angle ($V_{ub}=\vert V_{ub}\vert e^{-i\gamma}$)
 \begin{equation}
 \delta_{EW} = -\frac{3}{2}\frac{\vert V_{tb}^*V_{ts}\vert}{\vert V_{ub}^*V_{us}\vert}\frac{C_9+C_{10}}{C_1+C_2} = -\frac{3}{2}\frac{C_9+C_{10}}{C_1+C_2}\frac{\cot\theta_C}{\vert V_{ub}/V_{cb}\vert} \label{deltaewdefn}
 \end{equation}
In the above equation, $\theta_C$ is the Cabbibo angle. It has been argued \cite{neubert} that the
above combination of Wilson coefficients in $\delta_{EW}$ is renormalization group invariant to a good accuracy. From the above expression it is clear that the numerical value of $\delta_{EW}$ sensitively
depends on $\vert V_{ub}\vert$ (and to some extent on $V_{cb}$ also). 
At present, $\vert V_{ub}\vert^{excl}$ is very different from
$\vert V_{ub}\vert^{incl}$ \cite{pdg}.
 To incorporate both the ranges, we take the upper limit as extracted from the inclusive value and the lower limit as suggested by exclusive measurements. Therefore, within SM we have (typical central value
employed in literature is $\delta_{EW}\sim 0.64$)
\begin{equation}
 0.35 < \delta_{EW}^{SM} < 0.79
\end{equation}
Neglecting the amplitude $a$ and using the above expressions leads to
\begin{eqnarray}
 R_c &=& 1 - 2r_c\cos\delta_c(\cos\gamma - \delta_{EW}) \nonumber \\
&+& r_c^2(1 - 2\delta_{EW}\cos\gamma + \delta_{EW}^2)
\label{Rcdefn}
\end{eqnarray}
where $r_c=\frac{\vert T+C\vert}{\vert p\vert}$ and $\delta_c$ is the strong phase
difference between the amplitudes $(T+C)$ and $p$. Making further use of flavour $SU(3)$ symmetry,
and assuming factorization, it is possible to relate the magnitude of $(T+C)$ to the tree dominated decay $B^+\to\pi^+\pi^0$ \cite{grl}. Since amplitude $p$ can be extracted from $B^+\to K^0\pi^+$
rate, one arrives at
\begin{equation}
 r_c = \zeta_{SU(3)}\sqrt{2}\sqrt{\frac{BR(B^+\to\pi^+\pi^0)}{BR(B^+\to K^0\pi^+)}}
\end{equation}
where \[
\zeta_{SU(3)} = \frac{\vert V_{us}\vert}{\vert V_{ud}\vert}\frac{f_KF^{B\to\pi}(m_K^2)}{f_{\pi}F^{B\to\pi}(m_{\pi}^2)}
\frac{\lambda^{1/2}(m_B^2,m_{\pi}^2,m_K^2)}{
\lambda^{1/2}(m_B^2,m_{\pi}^2,m_{\pi}^2))}
             \]
 encodes the
$SU(3)$ breaking corrections within the approximations used. $F^{B\to\pi}$ is the $B\to\pi$ form-factor
and the function $\lambda(x,y,z)=x^2+y^2+z^2-2xy-2xz-2yz$ is the phase space factor. Using HFAG
values for the branching ratios, we arrive at $r_c=0.16\pm 0.04$, where we have used the 
form-factors as in \cite{ball} and errors have been added in quadrature (we have slightly inflated the
total error to account for departure from factorization assumption).
The authors in \cite{kpluspi0sumrule} arrive at the following expression after eliminating $\delta_c$,
and retaining terms to linear order in $r_c$,
\begin{equation}
 \left(\frac{R_c-1}{\cos\gamma -\delta_{EW}}\right)^2 + 
\left(\frac{A_{CP}(K^+\pi^0)}{\sin\gamma}\right)^2 = 4r_c^2 + {\mathcal{O}}(r_c^3)
\end{equation}
and conclude that due to the EW penguin contribution, the first term itself can saturate the
above sum-rule. Note that the authors have used a different value for $r_c$ compared to what is quoted 
above. This difference is essentially due to the form factors and phase space contributions not included in 
\cite{kpluspi0sumrule}.

We follow a slightly different route here. Instead of trying to check whether the sum-rule is
satisfied within SM, we invert the above relation and express $\delta_{EW}$ in terms of quantities
which are either directly measured, like rates and asymmetries,
 or are expressible in terms of measured quantities like $r_c$. 

%
\begin{figure}[ht!]
\hbox{\hspace{0.03cm}
\hbox{\includegraphics[scale=0.815]{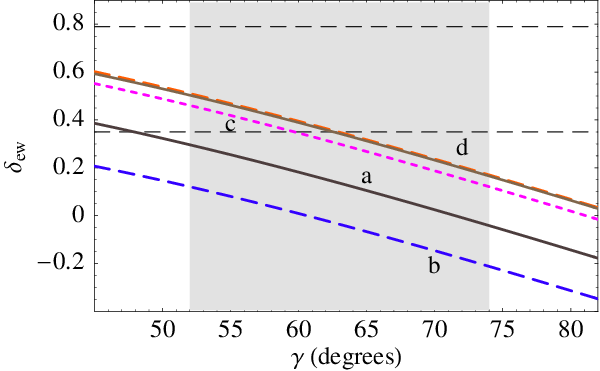}}
}
\caption{$\delta_{EW}$ as a function of $\gamma$. The horizontal dashed lines show the SM range for $\delta_{EW}$ while the vertical limits of the shaded region are the $2\sigma$ allowed range of $\gamma$.
 }
 \label{fig1}
\end{figure}
In Fig.\ref{fig1}, we plot
$\delta_{EW}$ as a function of CKM angle $\gamma$ (the shaded region represents
$2\sigma$ allowed range for $\gamma$ \cite{UTFIT, CKMfitter}) for different values 
of $R_c$, $A_{CP}(K^+\pi^0)$
and $r_c$. The horizontal dashed lines show the SM range of $\delta_{EW}$ quoted above.
Curves $a$, $b$ and $c$ refer to central, minimum and maximum values ($1\sigma$) for the above parameters while
$d$ shows two curves almost indistinguishable where a maximum-minimum combination of the parameters
is used. From the plot it is very clear that the curve corresponding to  central values of 
various quantities, curve $a$, always stays outside of the large SM range of $\delta_{EW}$. 
Curves $c$ and $d$ on the other hand fall within the SM range, albeit for smaller values of $\gamma$. 

\begin{figure}[ht!]
\hbox{\hspace{0.03cm}
\hbox{\includegraphics[scale=0.7805]{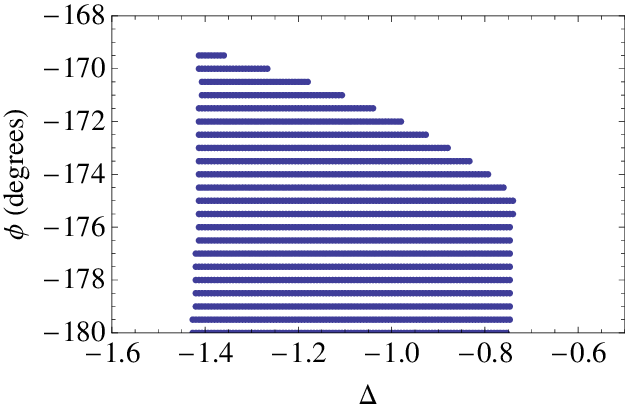}}}
\caption{Parameter space in $\Delta$-$\phi$ plane for $\delta_c=-20^{\circ}$
keeping other parameters fixed at their central values. 
 }
 \label{fig2}
\end{figure}

\begin{figure}[ht!]
\hbox{\hspace{0.03cm}
\hbox{\includegraphics[scale=0.815]{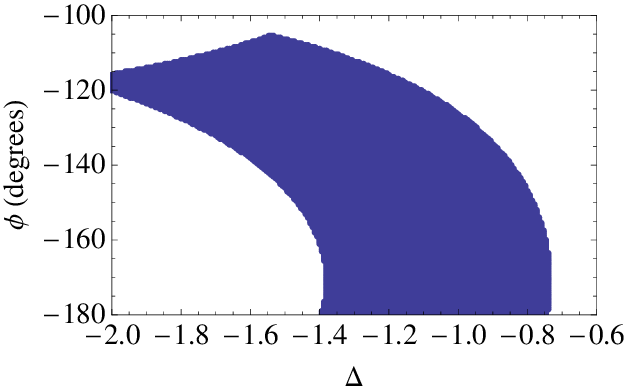}}}
\caption{Same as in Fig.(\ref{fig2}) but with $\delta_c=-10^{\circ}$
 }
 \label{fig3}
\end{figure}

Since the default curve, curve $a$, stays outside SM range of $\delta_{EW}$,
it is instructive to examine the impact of NP. This will be the case, if for example $\vert V_{ub}\vert$ eventually turns out to be close to the current inclusive value.
It is straightforward to incorporate the effects of NP to EW penguins by replacing $\delta_{EW}$ by
$\delta_{EW}\Delta e^{i\phi}$, where $\Delta$ denotes deviation from SM scaled by SM value and $\phi$ is the CP violating phase carried by the NP operators.
With this replacement, we arrive at
\begin{eqnarray}
 R_c^{new} &=& 1 - 2r_c\cos\delta_c (\cos\gamma - \delta_{EW}\Delta\cos\phi) \\
&+& r_c^2(1 + \delta_{EW}^2\Delta^2 + 2\delta_{EW}\Delta[-\cos\gamma\cos\phi +
\sin\gamma\sin\phi]) \nonumber
\end{eqnarray}
and
\begin{equation}
 A_{CP}^{new}(K^+\pi^0) = -\frac{2r_c}{R_c^{new}}\sin\delta_c (\sin\gamma + \delta_{EW}\Delta\sin\phi)
\end{equation}
In Fig.(\ref{fig2}) and Fig.(\ref{fig3}) we show the allowed range of
$\Delta$ and $\phi$ 
 for $\delta_c=-20^{\circ}$ and $-10^{\circ}$  respectively (no solution is found for the range of parameters employed for positive
$\delta_c=10(20)^{\circ}$). We have chosen $1\sigma$ experimental range for 
the observables $R_c$ and $A_{CP}(K^+\pi^0)$.
All other parameters are held to their central values and $\delta_{EW}=0.64$. We have varied $\Delta$ between $-2$ to $2$ and $\phi$ between $-\pi$ and $\pi$.
We find that the new CP violating phase $\phi$ and the magnitude $\Delta$ should be very large and negative (discrete ambiguities have been ignored at this point). This is consistent with the observation made in
\cite{fleischerzupan} in the context of CP asymmetry in $B\to K_S\pi^0$.
If this kind of a scenario turns out to be true, it will be a clear indication beyond the minimal flavour violation (MFV) hypothesis where the only source of CP (and flavour) violation is the CKM phase.


Given the present errors on various quantities, currently it is not very conclusive to infer
physics beyond SM from $B\to K\pi$ rates and asymmetries. As is clear from the above discussion,
$\vert V_{ub}\vert$ plays the most crucial role in reaching the conclusion whether there is
a sign of new physics or not. We have merged the two measurements (inclusive and exclusive determinations of $V_{ub}$) and shown a broader band, within which
one can accommodate the SM prediction. In this context, it is important to keep in mind that
the amplitude relations have been obtained after neglecting small contributions. One can ask
if inclusion of those neglected pieces can improve the situation. Very recently, the authors in \cite{ciuchinietal}, following the operator language, have included (some of the) $1/m_B$ contributions. Employing QCDF
\cite{bbns},
the authors have checked for the consistency of the fits to all observables in the 
$B\to K\pi$ system, as well as by removing
the one under consideration. Their results show that including the $1/m_B$ corrections
and varying them within very plausible ranges, the rates and CP asymmetries obtained are well in
agreement with the HFAG values. They conclude therefore that the inclusion of these formally $1/m_B$ suppressed terms 
(having a very marginal impact on branching ratios) naturally leads to opposite signs for $A_{CP}(K^+\pi^-)$
and $A_{CP}(K^+\pi^0)$, thereby making the SM prediction of $\Delta A_{K\pi}$ consistent with
experiments. Similar conclusions have been reached within PQCD \cite{li} and global fits based on (approximate) 
flavour $SU(3)$ symmetry \cite{su3global} where a large colour suppressed tree contribution is needed.
Inclusion of some of the neglected small contributions can be effectively seen as modifying
the value of $\delta_{EW}$, and can thus bring the data and theory in better agreement. This will then be consistent with the findings of \cite{ciuchinietal}.

From this discussion, it is clear that no significant new physics contribution may be needed
to address the $B\to K\pi$ puzzle(s). However, it is not just $B\to K\pi$ rates and asymmetries
that show a possible tension with SM expectations. A somewhat lower
value of $\Delta m_s$, large CP violating phase in the $B_s$ mixing,
polarization puzzles in $B\to VV(T)$, $\sin2\beta$ from penguin dominated modes
(see \cite{HFAG} for the present experimental status of all these), all call
for a closer look and have been advocated as hints of physics beyond SM. The latest Belle measurement \cite{belletaunutau}
of $BR(B\to\tau\nu_{\tau})$ confirms the larger than SM value reported earlier
\cite{taunutauold}. The largest
uncertainty comes from $f_B$ and $V_{ub}$. One can instead look at the ratio 
$BR(B\to\tau\nu_{\tau})/\Delta m_d$, which within SM takes the form
\begin{eqnarray}
 \frac{BR(B\to\tau\nu_{\tau})}{\Delta m_d} &=& \frac{3\pi}{4}\frac{m_{\tau}^2\tau_{B^+}}
{M_W^2 S(x_t)\eta_{B_d}B_{B_d}}\frac{1}{\vert V_{ud}\vert^2} \nonumber \\ 
&\times&\left(1-\frac{m_{\tau}^2}{m_{B^+}^2}\right)^2 
\left(\frac{\sin\beta}{\sin\gamma}\right)^2 
\end{eqnarray}
where $\eta_{B_d}=0.56$ is the QCD correction factor entering $\Delta m_d$ while $S(x_t)$ encodes the
dominant short distance top-quark contribution to the box diagram. This ratio brings out $\sim 2\sigma$
tension between the Lattice values of $B_{B_d}$ and that required by global fits to data \cite{ckmtalk}.
Independent of $BR(B\to\tau\nu_{\tau})$, employing the same lattice value of $B_{B_d}$ yields a consistent
value for $\Delta m_d$ for typical $f_B$ quoted in literature. A large central value (currently the errors are also large) of $BR(B\to\tau\nu_{\tau})$ as observed can not be easily accommodated
within SM, and definitely calls for new physics. 
Various new physics scenarios have been advocated in the literature in order to resolve $B\to K\pi$
and other $b\to s$ penguin dominated issues. A simple example is the sequential four generation standard model, SM4, which can simultaneously explain
$B\to K\pi$ puzzles and $B_s$ mixing while at the same time being consistent with other measurements \cite{Sm4}.
However, without invoking extra operator structures like scalar, tensor or right handed operators, it is not feasible to explain polarization puzzle in $B\to VV$ modes \cite{btovv}. Therefore, models with only
enhanced penguins can not simultaneously explain all these puzzles.
Looking at all these hints, we find it
very plausible that an extended Higgs sector is in fact needed. A general two Higgs doublet model (g2HDM),
allowing for CP violation can in fact resolve most of the above
mentioned discrepancies. The CP violating phase in the Higgs sector will be common to
Higgs (di-)penguin diagrams and therefore will have no effect on $B_d$ mixing while
making a non-negligible contribution to $B_s$ mixing, due to non-zero strange quark mass as opposed to negligible down quark mass. For the same reason,
$b\to s\bar{s}s$ penguin processes will receive an additional contribution compared to $b\to s\bar{d}d (\bar{u}u)$. Therefore,
there is a possibility of resolving polarization puzzle (via the scalar-pseudoscalar operators in such a model \cite{londonvv}) that shows up in $b\to s$ penguin dominated
modes ($B\to VT$ is an exception and may have to do with a very different hadronic structure of the tensor meson involved). In such a model, the strength of EW penguin operators will also get modified,
possibly bridging any gap between theoretical predictions and experimental measurements. One does not expect these modifications to be numerically very large once the latest $b\to s\gamma$ constraints are taken
into account. Very roughly speaking, $b\to s\gamma$ rate is practically independent of $\tan\beta$ 
for $\tan\beta > 2$ \cite{misiaketal}, and only places tight constraints on $m_{H^+}$ which can be combined
with $B\to\tau\nu_{\tau}$ measurements to eliminate a large region of parameter space. Further immediate constraints come from limits on $B_s\to\mu^+\mu^-$ branching fraction. 
 A detailed phenomenological study in the context of a specific 2HDM will be presented
elsewhere.

In this note, using flavour $SU(3)$ symmetry, we have shown that given the present errors on various
quantities, mainly $V_{ub}$, rates and asymmetries in $B\to K\pi$ modes do not show any significant deviation from SM expectations. Inclusion of formally suppressed contributions will not significantly affect the rates but can have a large impact on CP asymmetries and can bring theory and data in better agreement. We have also found that if EW penguins are to resolve the discrepancies in the
$K\pi$ modes, a large new CP violating phase is needed. Looking at various other observables, a general two Higgs doublet model may offer the simplest resolution to most of the puzzling issues in flavour physics. However, we strongly feel that a precise measurement of $V_{ub}$ and lattice estimation of $f_{B_{d,s}}$ and $B_{B_{d,s}}$ is immediately needed, without which possible NP may remain
hidden under the parametric uncertainties.




%

\end{document}